\documentclass[aps,prd,twocolumn,groupedaddress]{revtex4-2}

\usepackage{graphicx}
\usepackage[dvipsnames]{xcolor}
\usepackage{footnote}
\usepackage{hyperref} 
\usepackage[left]{lineno}
\def\ve{\varepsilon}
\begin{document}
\title{Intergalactic magnetic field studies by means of \texorpdfstring{$\gamma$}{gamma}-ray  emission from GRB 190114C}

\author{P. Da Vela$^{1,2}$}
\email[]{Paolo.Da-Vela@uibk.ac.at}
\author{G. Mart\'{i}-Devesa$^1$}
\author{F.G. Saturni$^{3,4}$}
\author{P. Veres$^{5,6}$}
\author{A. Stamerra$^3$}
\author{F. Longo$^{7,8}$}
\affiliation{$^1$ Institut f\"{u}r Astro- und Teilchenphysik, Universit\"{a}t Innsbruck, A-6020 Innsbruck, Austria}
\affiliation{$^2$ INFN, Sezione di Pisa, Largo B. Pontecorvo 3, 56127 Pisa, Italy}
\affiliation{$^3$ INAF - Osservatorio Astronomico di Roma, Via Frascati 33, I-00078 Monte Porzio Catone (RM), Italy}
\affiliation{$^4$ ASI - Space Science Data Center, Via del Politecnico snc, I-00133 Roma, Italy}
\affiliation{$^5$ Department of Space Science, University of Alabama in Huntsville, Huntsville, AL 35899, USA}
\affiliation{$^6$ Center for Space Plasma and Aeronomic Research (CSPAR), University of Alabama in Huntsville, Huntsville, AL 35899, USA}
\affiliation{$^7$ Istituto Nazionale di Fisica Nucleare, Sezione di Trieste, I-34127 Trieste, Italy}
\affiliation{$^8$ Dipartimento di Fisica, Università di Trieste, I-34127 Trieste, Italy}

\date{\today}
\begin{abstract}
The presence of delayed GeV emission after a strong transient, such as a GRB (Gamma-Ray Burst), in the VHE (Very-High Energy, $E>100$ GeV) band can be the signature of a non-zero magnetic field in the intergalactic medium. We used a synchrotron self-Compton multiwavelength model to infer an analytical description of the intrinsic VHE spectrum (corrected for absorption by the Extragalactic Background Light, EBL) of GRB~190114C to predict the lightcurves and SEDs of the delayed emission with Monte Carlo simulations for different IGMF (Intergalactic Magnetic Field) configurations (strengths $B=8\times10^{-21}$ G, $10^{-20}$ G, $3\times 10^{-20}$G and correlation length $\lambda>1$ Mpc), and compared them with the \emph{Fermi}-LAT (\textit{Fermi} Large Area Telescope) limits computed for several exposure times. We found that \emph{Fermi} LAT is not sensitive enough to constrain any IGMF strengths using GRB 190114C.   
\end{abstract}
\maketitle
\section{Introduction \label{sect:intro}}

Magnetic fields are present everywhere in the Universe, from stars to galaxies and even clusters of galaxies. But the origin of the large-scale magnetic fields 
is one of the long-standing problems in cosmology.  There is a general agreement that the magnetic fields in the galaxies originate from the amplification of pre-existing weak seed fields (see e.g \cite{Widrow02} and \cite{Kulsrud08}). However, the origin of these seeds is still not known. Two main hypotheses exist: the astrophysical scenario and the cosmological scenario (see e.g \cite{Grasso01} and \cite{Durrer13}). If the magnetic fields originate in the early Universe, then a non-zero magnetic field is expected in the Intergalactic Medium (IGM) today. Whereas if the magnetic fields originate in large-scale structures during their formation, a negligible Intergalactic Magnetic Field (IGMF) would be expected unless galactic outflows effectively seed the magnetic fields in the deep IGM. Recently, Jedmazik \& Pogosian \cite{Jedamzik20} showed that the presence of primordial magnetic fields originated before recombination could resolve the discrepancy between the measurement of the Hubble constant derived by the Planck Collaboration \cite{Planck20} and the one performed by means of type Ia supernovae \cite{Reid19}. To shed some light on the origin of the magnetic fields  it is crucial to look for signatures of magnetization in the voids among the galaxies. Due to the difficulties of direct detection (e.g. \cite{Pshirkov16}), the observation of extragalactic $\gamma$-ray sources can be used to constrain the IGMF.        

Very-High Energy (VHE, $E>100$ GeV) gamma-rays from extragalactic sources are not able to propagate over large distances ($\sim 1$ Gpc) because they are absorbed by the Extragalactic Background Light (EBL) via the pair-production process ($\gamma+\gamma\rightarrow e^++e^-$) \cite{Nikishov62, Gould66}. For this reason, the primary VHE spectra of the sources are partially absorbed during the propagation in the IGM. 
The larger the distance of the source, the more pronounced is this effect. 

In addition, the EBL absorption is stronger for higher primary photon energies. The created pairs lose energy by means of the Inverse Compton (IC) process with the Cosmic Microwave Background (CMB) producing secondary $\gamma$-rays. Typical energies of the IC photons are $E\simeq70(E_0/10\textrm{ TeV})^2$ GeV \cite{Neronov09}, where $E_0$ is the energy of the primary source photon. Yet a non-negligible IGMF can deflect the pairs during their propagation to Earth. Due to the subsequent longer path length, the secondary GeV $\gamma$-rays result in a "pair-echo" delayed with respect to the primary emission from the source. The presence of this new component in the GeV domain provides a way to study the IGMF. This method was first proposed by Plaga \cite{Plaga95} and later developed by Ichicki et al. \cite{Ichiki08}, Murase et al. \cite{Murase08}, and Takahashi et al. \cite{Takahashi11} in the context of Gamma-ray Bursts (GRBs).

GRBs have been proposed to derive limits on IGMF (see e.g. \cite{Veres17}), and the recent discovery of VHE emission from GRB 190114C \cite{MAGIC19} (redshift $z\simeq0.42$) was used to constrain the IGMF. 

Wang et al. \cite{Wang20} performed an analytical calculation of the echo emission flux for different IGMF strengths and observing times. For their calculation they assumed a power law with spectral index 2, which is slightly harder than the 2.22 index reported by the MAGIC Collaboration between 200 GeV and 1 TeV as the primary source spectrum. The flux was then extrapolated up to 6 s after GRB trigger time which is where, reasonably, the afterglow emission started \cite{Ravasio19}. Comparing the predicted pair-echo Spectral Energy Distributions (SEDs) with the \textit{Fermi} Large Area Telescope (\textit{Fermi}-LAT) upper limits, the authors derived a lower bound on IGMF $B>10^{-19.5}$ G assuming a correlation length $\lambda\leq 1$ Mpc. They also verified that changing the maximum energy of the primary spectrum from 1 TeV to 15 TeV does not affect the result. On the other hand, Dzhatdoev et al. \cite{Dzhatdoev20} first reconstructed the primary source spectrum from the VHE spectral data points testing several EBL models and looking for a possible cutoff at higher energies. Then they used the publicly available code ELMAG3 \cite{Elmag3} to predict the pair echo emission from 20000 s after the burst time to 1 month. 
The VHE flux used by the authors, in this case, is the one measured by the MAGIC Collaboration during the time window 62 s -- 2400 s after the GRB trigger time. Comparing the predicted pair-echo SED with the \textit{Fermi}-LAT upper limits in the GeV domain, the authors conclude that the sensitivity of the \textit{Fermi} LAT is not sufficient to constrain the IGMF.

In this paper, we present the calculation of the expected pair echo SED and lightcurve for several observation times and IGMF strengths using a different approach. The choice of the GRB intrinsic spectrum is a key point: differently from \cite{Wang20} and \cite{Dzhatdoev20} we do not use a purely phenomenological primary spectrum, but a physically motivated synchrotron self-Compton (SSC) spectrum fitting the multiwavelength observations of the GRB afterglow. 
Then we used \mbox{CRPropa 3} \cite{crpropa3} to simulate the cascade emission in the GeV domain and derive the SEDs and lightcurves for several IGMF strengths and observation times, taking into account the time activity of the GRB in the VHE band. Finally, we compared the simulated lightcurves and SED with the results obtained by analyzing the \textit{Fermi}-LAT data.

\section{Analytic description \label{sect:analytic}}

To identify the relevant aspects required in our simulation, we begin with the analytic description of the involved processes. The flux produced by the cascade radiation is given by IC \cite{Dermer11} between the electron-positron pairs and the CMB assuming Thomson scattering:

\begin{equation}\label{eqn:flux}
    f_{\varepsilon_s} = \frac{3}{2} \left(\frac{\ve_s}{\ve_0}\right)^2 \int \frac{d\gamma}{\gamma^{4}} \left(1-\frac{\ve_s}{4\gamma^2\ve_0}\right) \int d\gamma_i C_T \frac{f_\ve (e^{\tau_{EBL}} -1)}{\ve^2}
\end{equation}

where $f_{\ve_s}=E_\gamma^2 dN/dE_\gamma$ is the scattered $\nu F_\nu$ flux at energy $\ve_s$ measured in units of $m_e c^2$, $\ve=E_\gamma/m_ec^2$ is the energy of the VHE photons directly produced by the GRB, $\gamma_i=\ve/2$ is the Lorentz factor of the pairs, $\tau_{EBL}$ is the optical depth of the EBL.
The inner integral describes the production of pairs by the VHE spectrum ($f_\ve=E^2F_E^{GRB}$).
The outer integral accounts for the IC scattering of the pairs on the CMB with typical energy $\ve_0=2.7 kT_{ CMB}/m_e c^2\approx 1.24\times 10^{-9}$. Note that Eq. \ref{eqn:flux} only accounts for the first generation of the cascade, and the pairs will only radiate for a time $\Delta T_{ IC}=\lambda_T/2\gamma c$. Here $\lambda_T=3m_e c/4 \sigma_T u_0 \gamma$  is the IC cooling length of a pair in the CMB with $u_0$ energy density.

We can account for the finite duration ($\Delta T_{activity}$) of the VHE emission and for the finite observation $\Delta T_{ obs}$ window of the \textit{Fermi} LAT by scaling  the expression for $f_{\ve_s}$ by the ratio of these timescales, $C_T$. The photons that contribute to the echo flux need to arrive in the window defined by the observation time, the angular spreading time, $\Delta T_{\rm A}=(\lambda_T+\lambda_{\gamma \gamma})/2\gamma^2 c$ and the echo duration from the deflection in the IGMF $\Delta T_{ B}=(\lambda_T+\lambda_{\gamma \gamma})\theta_B^2/2c$, where $\theta _B$ is the pair deflection angle induced by the IGMF. Here $\lambda_{\gamma\gamma}=D/\tau_{EBL}$ is the mean free path of the VHE photons before interacting with the EBL for a source at distance $D$. 

The delay of an echo photon compared to the photon arriving directly, without undergoing absorption is determined by a simple geometry \cite{Neronov09}, 

\begin{equation}
    c\Delta t=\lambda_{\gamma \gamma}+x-D\approx \frac{\lambda_{\gamma\gamma}}{2}\theta_B^2\left(1-\frac{\lambda_{\gamma\gamma}}{D}\right) \rm{,}
\end{equation}

where $x$ is distance travelled by the IC cascade photons. In the case of simulations, we also know the arrival times of individual photons and we account for different emission and observation scenarios by considering the arrival times of individual simulated photons.

\section{Simulation of pair-echo emission \label{sect:crpropa}}

In order to model the pair-echo emission for different IGMF settings we used the Monte-Carlo code \mbox{CRPropa} \cite{crpropa3}: given a particular primary photon spectrum this code traces the development of the cascade in the IGM. Hereafter we assume the cosmological parameters $H_0=70$ km s$^{-1}$ Mpc$^{-1}$, $\Omega _{\Lambda} = 0.7$, and $\Omega _{M} = 0.3$. The source is located at the centre of a sphere of radius $D$, which corresponds to the co-moving distance of the Earth to GRB 190114C ($z=0.42$). In order to contain a standard GRB jet aperture, we conservatively inject and track all primary photons within a $10^\circ$ cone and, as target photon fields for $\gamma$-$\gamma$ and IC interactions, we use the CMB and the Franceschini et al. \cite{Franceschini08} model for the EBL background. A photon that hits the sphere and has energy larger than 0.05 GeV represents a particle arriving and being detected at Earth. The magnetic field is assumed to be a turbulent zero-mean Gaussian random field with a Kolmogorov spectrum; it is defined in the Fourier space, transformed into real space, and then projected onto a 
(50 Mpc)$^3$ grid with $100^3$ cells. The minimum scale that can be resolved is 1 Mpc and the maximum set scale is 25 Mpc. For such a configuration the correlation length is $L_c\simeq5$ Mpc. Given the primary gamma ray photon energies used here (i.e. $0.2$--$10$ TeV), the correlation length is much larger than the loss length of the pairs (the largest loss length would be $\lambda_T=0.8$ Mpc for $E_{\gamma}=0.2$ TeV). In this regime, the deflection angle of the pairs does not depend on the correlation length. The eventual lower bound on the IGMF can be easily re-scaled for the low correlation length regime considering the dependence of the deflection angle on the correlation length, this is $\theta _B = \sqrt{\lambda_B \lambda_T}/R_L$, where $R_L$ is the Larmor radius of the pair \cite{Neronov09}. The grid is periodically repeated to cover the whole volume between the GRB and the Earth ($D\simeq 1.6$ Gpc). For each magnetic field strength (root mean square) tested, we used CRPropa to inject $10^3$ primary $\gamma$-ray photons and repeated the procedure $10^3$ times (i.e. simulating $10^6$ photons in total). Further, for each run we changed the seed used to generate the magnetic field grid in order to avoid spurious features due to the choice of that particular realization of a magnetic field. All particles are traced with a minimum step size of $10^{-4}$ pc, which is sufficient to reproduce time delay with an accuracy better than 3 hours. We only consider the first cascade generation, since we find the contribution of further generations to be negligible for these settings.

The choice of the primary spectrum to be injected in the IGM is a crucial point that highly impacts the derived cascade spectrum. Hence solely a realistic, physically motivated intrinsic spectrum will provide a sensible cascade flux. With this in mind, we inferred the VHE spectral shape at energies higher than 1 TeV from the SSC model fitted to the multiwavelength SED by the MAGIC Collaboration \cite{MAGIC19Model}. We estimated the best-fit parameters of the time-averaged log-parabola shape in the energy range $0.2$--$1$ TeV:
\begin{equation}\label{eqn:logpar}
    F_E^{GRB} \propto \left(
    \frac{E}{E_0}
    \right)^{-\langle \alpha \rangle - \langle \eta \rangle \log{\left(
    E/E_0
    \right)}}
\end{equation}
Here $E_0$ is the pivot energy, $\langle \alpha \rangle$ is the average spectral slope and $\langle \eta \rangle$ is the spectral curvature. First, we fixed $E_0$ at 0.4 TeV as done in \cite{MAGIC19Model}; then, we estimated $\langle \alpha \rangle = 2.51$ and $\langle \eta \rangle = 0.21$ by averaging the GRB 190114C spectral slopes and curvature indices in different time bins presented in their Table 1.

To build the SED of the cascade emission in the \textit{Fermi}-LAT band we first calculated the arrival directions of the cascade photons. This is needed because the SED is computed within the point spread function (PSF) of the \textit{Fermi} LAT. Following the scheme presented in \cite{Dermer11} (Fig. 1) the observer is assumed to be perfectly aligned with the emission cone axis: in such a configuration the cascade photon is detected at an angle $\theta$ with respect to the line of sight given by $\sin\theta=(\lambda_{\gamma\gamma}/D)\sin\theta_B$, where $\lambda_{\gamma\gamma}$ is the mean free path of the primary $\gamma$-ray photon, $D$ is the distance to the source. Considering $T_0=$ 20:57:03.19~UTC as the burst trigger time \cite{Gropp19}, 
we looked for the echo emission after $T_0+2 \cdot 10^4$ s to exclude all photons associated with the GRB afterglow in the GeV domain \cite{Fermi20GRB}. The cascade spectrum within a certain observation time $\Delta T$ is calculated this way:

\begin{displaymath}
    F_E=\frac{F^{GRB}(>200 \textrm{ GeV})}{F_{sim}}\frac{\Delta N_{cascade}(E, \theta<\theta_{PSF})}{\Delta T\Delta S \Delta E}=
\end{displaymath}
\begin{equation}\label{sim_cascade}
    \frac{F^{GRB}(>200 \textrm{ GeV})}{\Delta N_{sim}} \frac{\Delta T_{activity}}{\Delta T} \frac{\Delta N_{cascade}(E, \theta<\theta_{PSF})}{\Delta E}
\end{equation}

where $F^{GRB}(E>200 \textrm{ GeV})$ is the integrated flux (number of photons/$\textrm{cm}^2$ $\textrm{s}$) of the GRB measured in the VHE band, $F_{sim}$ is the integrated flux of the GRB inferred from the simulation in the same energy band, $\Delta N_{sim}$ is the total number of injected GRB photons not absorbed by the EBL for all realisations (i.e. after $10^3$ simulations), $\Delta S$ is the projected simulation area for our $10^{\circ} $ cone selection, $\Delta T_{activity}\simeq40$ minutes is the time activity of the GRB in the VHE band, $\Delta N_{cascade}$ is number of cascade photons collected at energy $E$ and within $\theta_{PSF}$, and $\theta_{PSF}$ is the \textit{Fermi} LAT's PSF 68\% containment angle at 1 GeV \cite{Ackermann12}. Concerning $F^{GRB}(E>200 \textrm{ GeV})$, MAGIC telescopes started to observe the GRB after at $T_0+62$ s. Since the VHE emission likely started at $T_0+6$ s (when the power law decay of the afterglow starts) we extrapolated the measured flux down to $T_0+6$ s using the best fit power law decay to the VHE data \cite{MAGIC19}. This provides a total flux about a factor of five larger than the average flux published by the MAGIC Collaboration. 

\begin{figure}[!tbh]
\begin{center}
\includegraphics[width=\hsize]{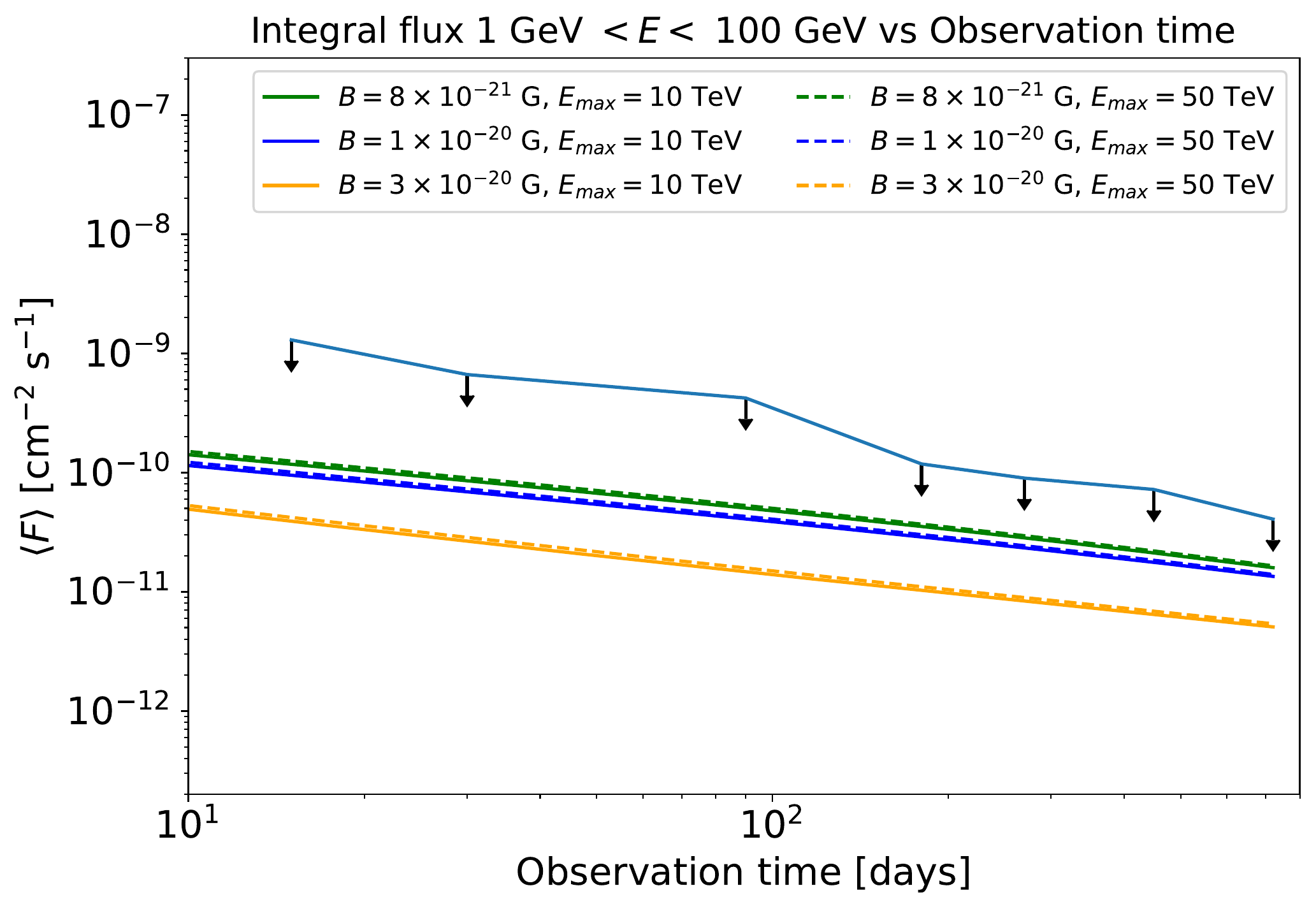}
\end{center}
\caption{\small Expected echo daily lightcurves between 1 GeV and 100 GeV for different IGMF strengths and maximum primary energies. The lightcurves are plotted together with the \textit{Fermi}-LAT upper limits.}
\label{lightcurve_logpar}
\end{figure}

The upper limits in the \textit{Fermi}-LAT band have been derived from $T_0+2 \cdot 10^4$ s for different exposure times (see next Sec. \ref{sect:fermi}). To take into account the dilution in time of the echo flux, both the spectra $F_E(E)$ and the lightcurves $F(T)$ have been averaged over the corresponding time window. Given a certain exposure time $T$ we then calculated $\langle F_E(T)\rangle =\int_0^T F_{E}(t)dt/T$ and $\langle F(T) \rangle =\int_0^T F(t)dt/T$ from the simulations. The lightcurves have been evaluated in the same energy range used to compute the upper limits in the GeV domain (1 GeV $<E<$ 100 GeV). In Fig. \ref{lightcurve_logpar} the expected lightcurves for different IGMF strengths are plotted together with the \textit{Fermi}-LAT upper limits derived for 15 days, 1, 3, 6, 9, 15 and 24 months of observation time. Concerning the magnetic field strengths, we tested the same as in \cite{Wang20} – this is, $B=10^{-20}$ G and $3\times 10^{-20}$ G. Since none of them can be constrained, we also tested a weaker strength, namely $B=8\times 10^{-21}$~G. Given that the flux does not change dramatically for even lower magnetic field strengths we decided not to decrease further the tested IGMF strength. As stated before, we conservatively set the maximum energy of the primary GRB to the one reported by the MAGIC Collaboration, namely 10 TeV. However, to test how this choice can affect our procedure we also tested $E_{max}=50$ TeV. The results are plotted in the same figure together with the case $E_{max}=10$ TeV.

\begin{figure*}[!tbh]
\begin{center}
\includegraphics[width=0.49\hsize]{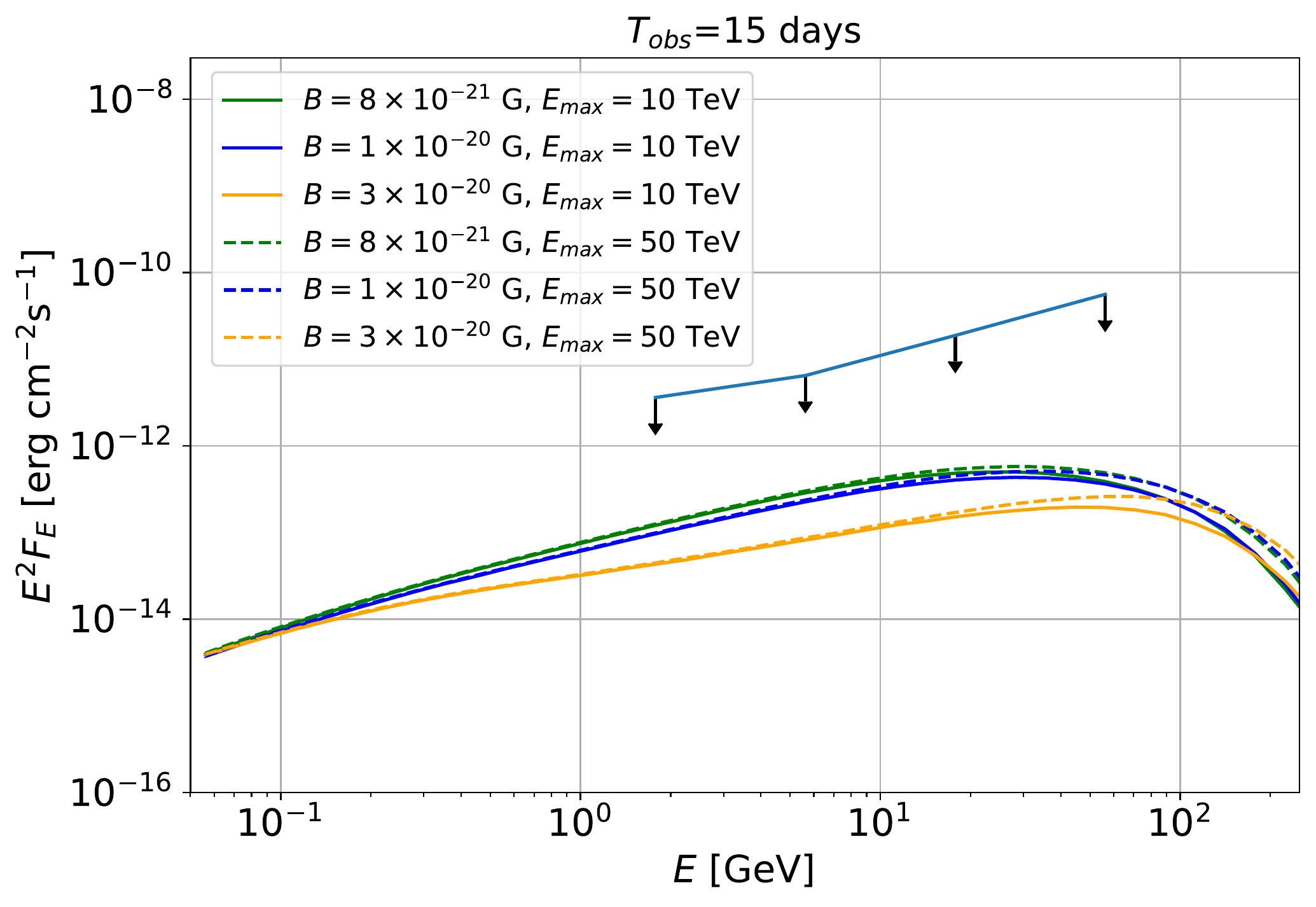}
\includegraphics[width=0.49\hsize]{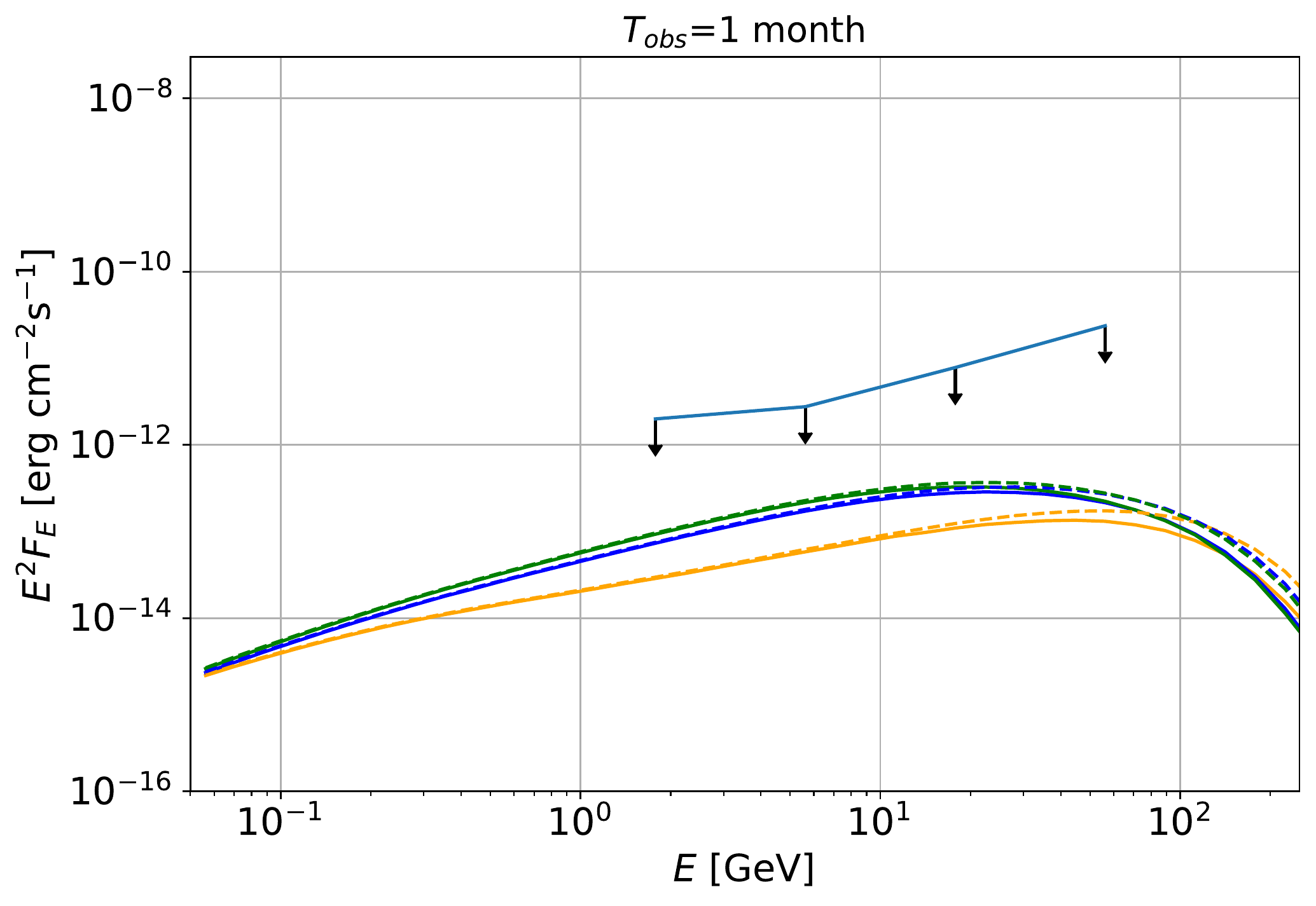}
\includegraphics[width=0.49\hsize]{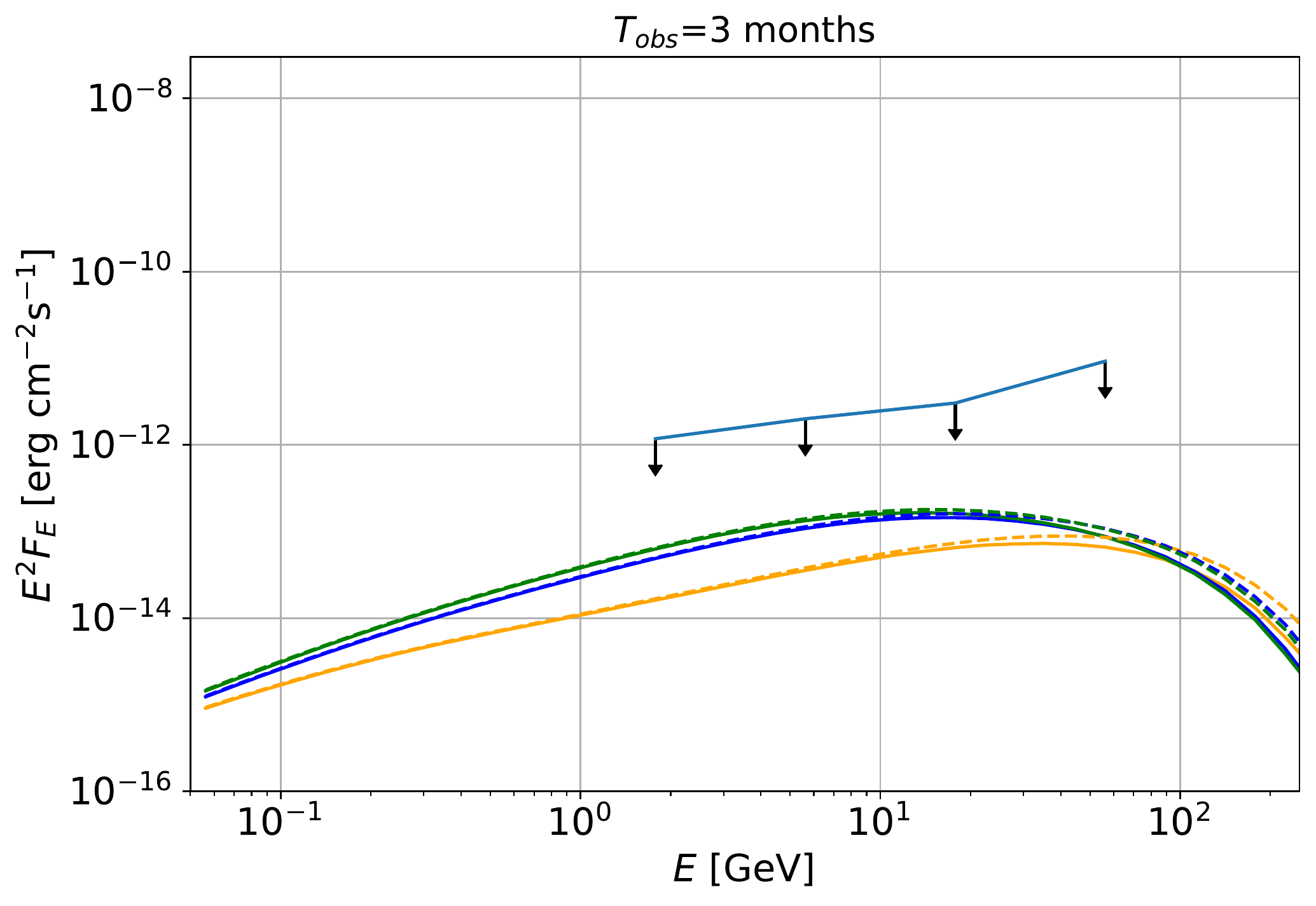}
\includegraphics[width=0.49\hsize]{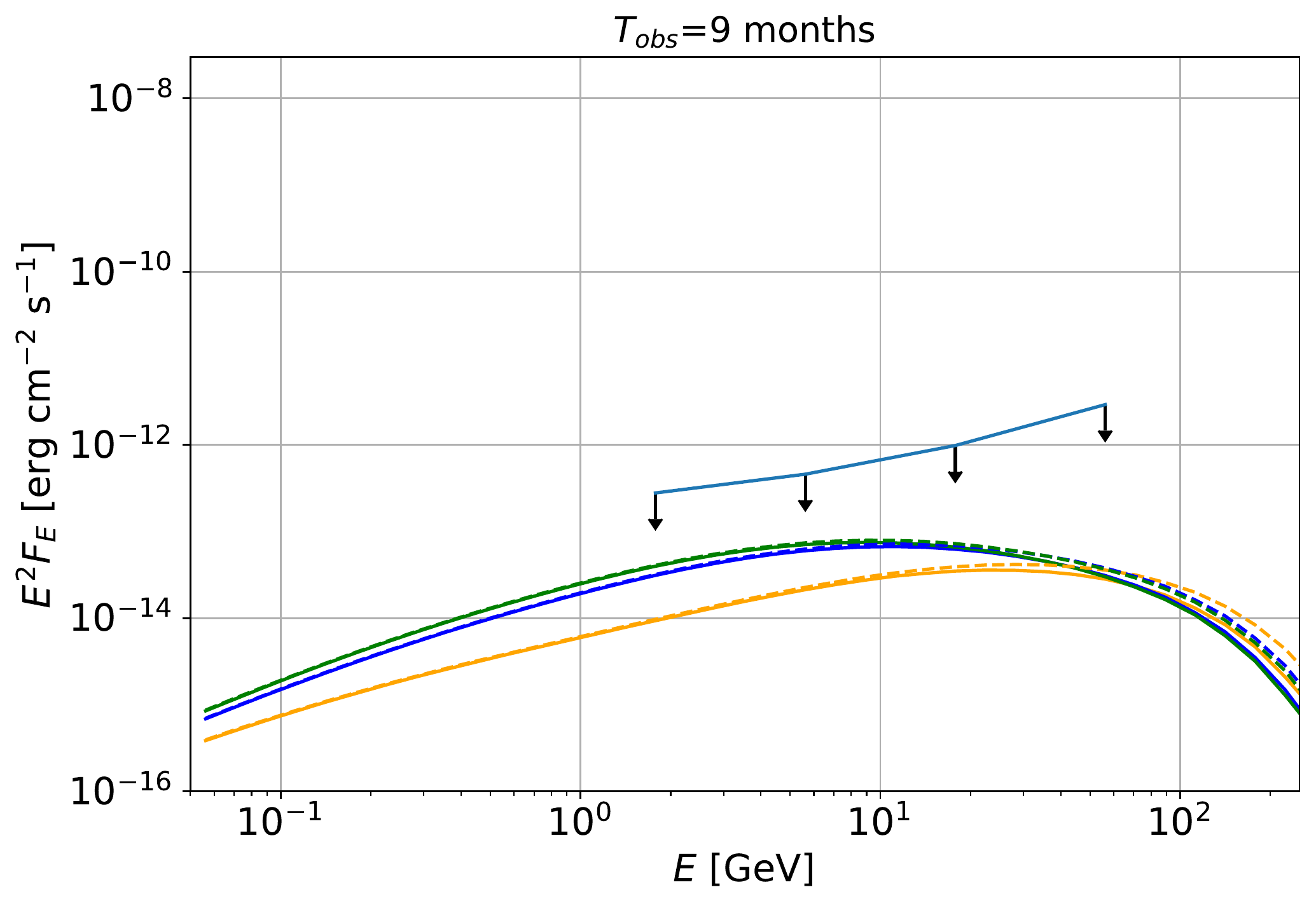}
\includegraphics[width=0.49\hsize]{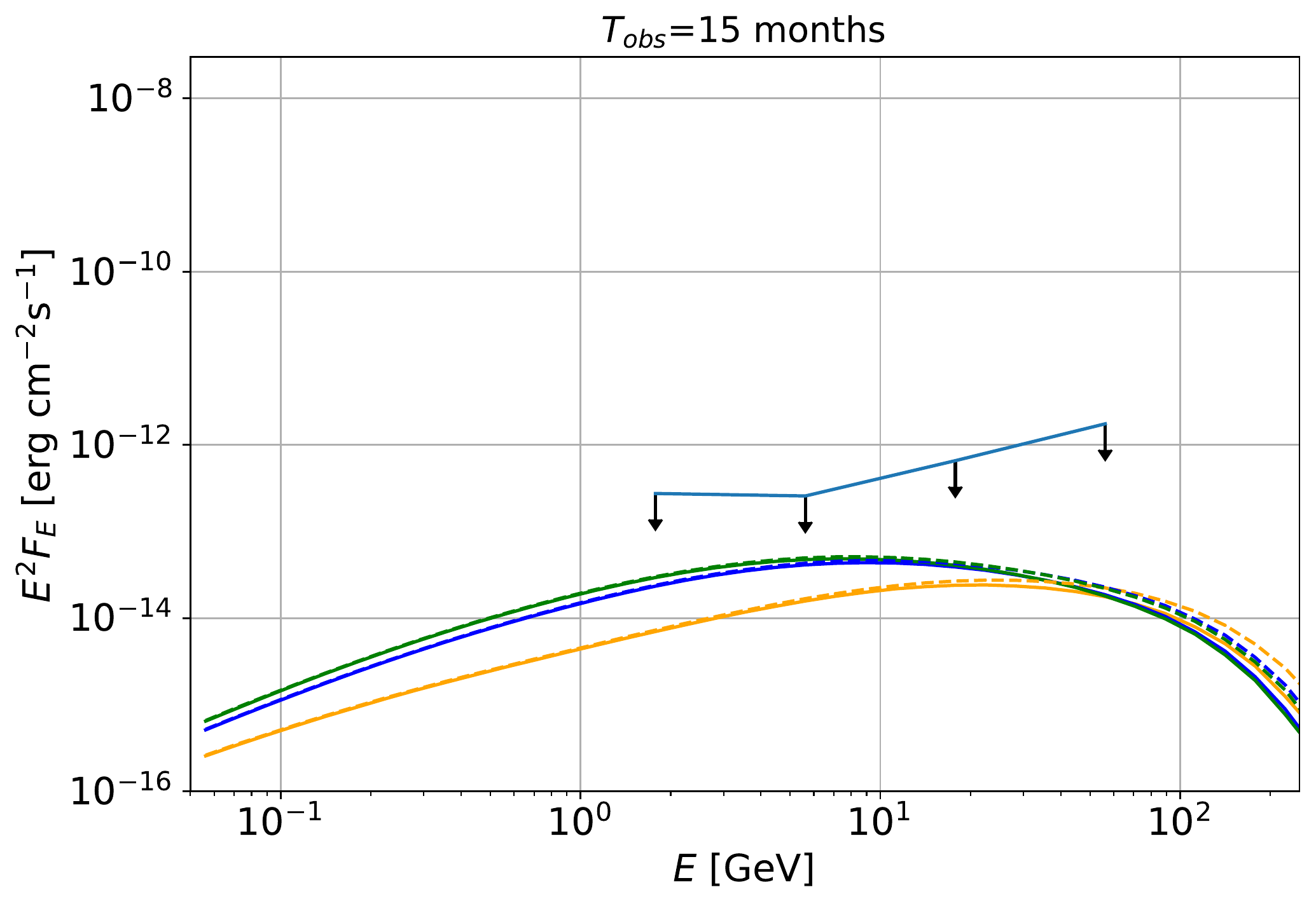}
\includegraphics[width=0.49\hsize]{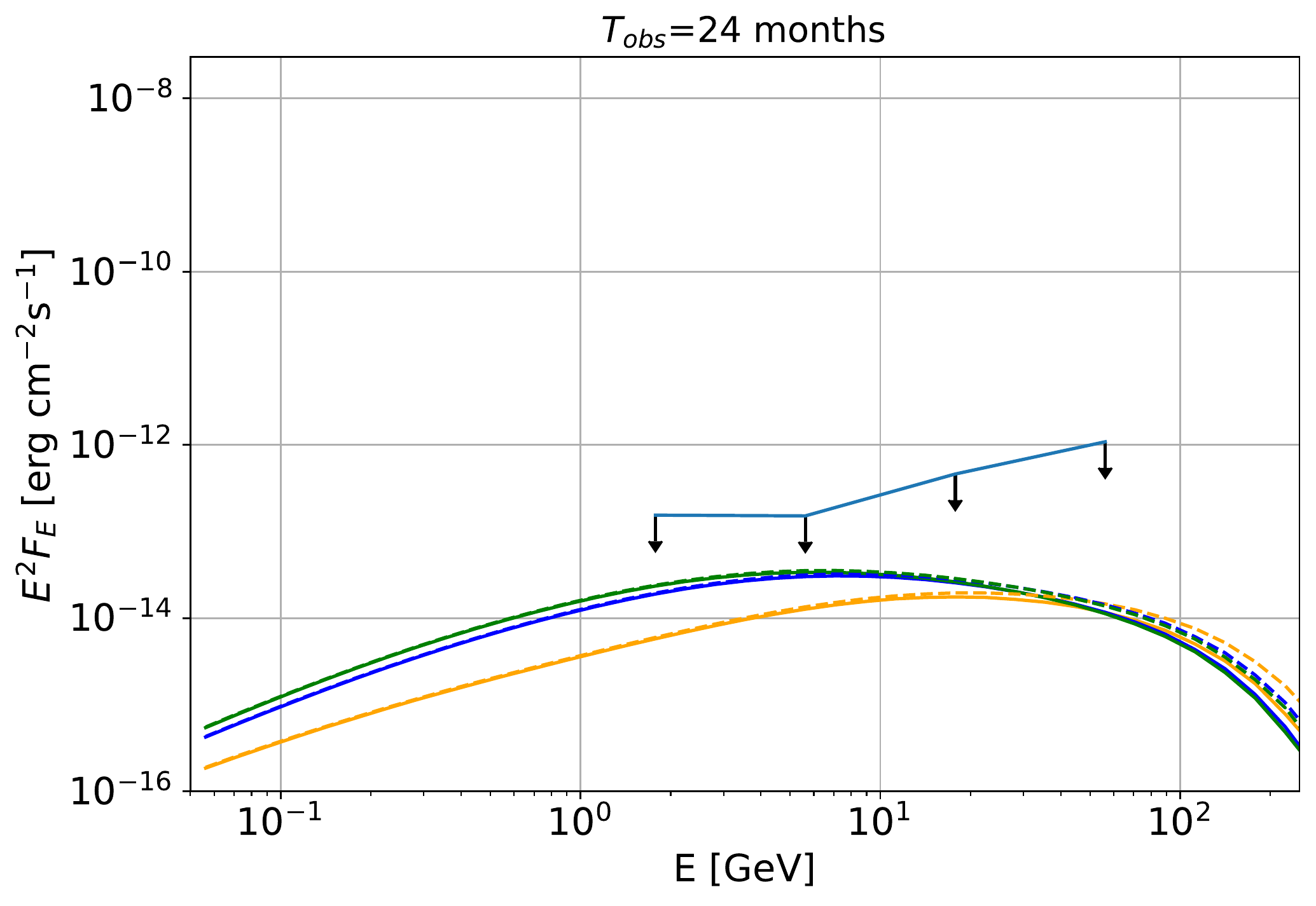}
\end{center}
\caption{\small Expected SEDs for different IGMF strengths, observation times and for $E_{max}=10$ and 50 TeV. The \textit{Fermi}-LAT differential upper limits are also shown.}
\label{seds_logpar}
\end{figure*}

In Fig. \ref{seds_logpar} we reported the expected SEDs ($E^2F_E$) as inferred from the simulations for different IGMF strengths, $E_{max}=10$ and 50 TeV and for different exposures. The SEDs are plotted together with the differential upper limits of the \textit{Fermi} LAT.

\section{\textit{Fermi}-LAT data analysis \label{sect:fermi}}

The simulations described before are compared with data from the \textit{Fermi} LAT \cite{Fermi}. We include observations taken between $T_0 + 2 \cdot 10^4$ s and $T_0 + 24$ months, selecting events with energies between 1 and 100 GeV in a region of interest (ROI) of $10^{\circ}\times 10^{\circ}$ centred on the GRB coordinates \cite{SwiftGRB19}. As previously stated, this selection guarantees no contamination from the burst itself \cite{Fermi20GRB} and focuses on the most sensitive energy range of the \textit{Fermi} LAT. We select \texttt{P8R3 SOURCE} data (evclass $=128$) with a \texttt{FRONT+BACK} event type (evtype $= 3$), applying a maximum zenith angle cut at $100^{\circ}$ to prevent Earth limb contamination.

Using \textit{Fermitools} (version 2.0.8) and \textit{fermipy} (version v1.0.1) \cite{Fermipy}, we perform a binned maximum likelihood analysis on our dataset \cite{Mattox96}. Subsequently, we account for the PSF and energy dispersion ($\rm{\texttt{edisp\_bins}}=-1$; excluding the isotropic diffuse component) using the instrument response functions \texttt{P8R3\_SOURCE\_V3}. As our background source model we use a $15^{\circ}\times 15^{\circ}$ selection of the 4FGL-DR2 ('gll\_psc\_v27') catalogue \cite{4FGL, 4FGL-DR2} centred on the burst together with the recommended galactic and isotropic diffuse components -- 'gll\_iem\_v07' and 'iso\_P8R3\_SOURCE\_V3\_v1', respectively. The detection significance of these ROI sources is evaluated with the test statistic $TS=-2\left( L_{0} / L_{1}\right)$, where $L_{0}$ is the log-likelihood of the null hypothesis and $L_{1}$ the log-likelihood of the complete model. After a preliminary iterative optimization of the ROI (\texttt{optimize} function, fitting first sources with larger predicted counts based on the catalogue), sources detected with $TS<4$ (i.e. $2\sigma$) are removed to avoid unnecessary degrees of freedom. We also notice that the blazar PKS 0346--27 is in our ROI and has been flaring occasionally since 2018, thus including the observational window of this study \cite{Angioni19, Gokus19, Mereu20}.  Its spectral model from the 4FG-DR2 catalogue -- a log-parabola -- does not characterise properly the flaring state, while a power law with an exponential cut-off can account for the spectrum observed by the \textit{Fermi} LAT. We therefore modify accordingly the background model and free the spectral parameters of PKS 0346--27 in our fit, together with the normalization of all sources within $3^{\circ}$ of the ROI's centre. Such analysis is performed in datasets lasting 0.5, 1, 3, 6, 9, 15, and 24 months with no detection ($TS$ lies between $0.0$ and $0.1$ for the different $\Delta T$), therefore we extracted upper limits at 95\% confidence level. We achieved this by adding a point source modelled as a power law with spectral index 2 at the GRB nominal position. No significant difference is found assuming the spectral shape of the putative cascade obtained from the simulations. \\

\section{Discussion and conclusions \label{sect:discussion}}

In this paper, we used the $\gamma$-ray emission from GRB 190114C to infer the pair echo SED and lightcurves for different IGMF strengths. We used CRPropa 3 to simulate the cascade emission in the GeV domain originated by the interaction of the primary VHE GRB spectrum with the IGM. We then compared the expected SEDs and lightcurves with the differential and integrated flux upper limits derived by analyzing the \textit{Fermi} data. 
From both Fig. \ref{lightcurve_logpar} and \ref{seds_logpar} we clearly see that no IGMF strengths can be constrained because the flux upper limits are well above the predicted cascade flux. For a given observation time, the amount of cascade flux depends on the strength of the IGMF: as expected, increasing the IGMF strength the cascade is more diluted in time due to the larger delay experienced by the pairs, and the largest tested magnetic field strength always corresponds to the lowest cascade flux (Fig. \ref{lightcurve_logpar} and Fig. \ref{seds_logpar}). This is also compatible with the results in \cite{Wang20} and \cite{Dzhatdoev20}. 

The evolution of the SEDs as a function of observation time and the shape of the lightcurves can be explained in this way: for $T_{obs}=15$ days we have the maximum level of cascade flux. On the other hand for such an exposure time the \textit{Fermi} limits are also the largest. As soon as we increase the observation time, the \textit{Fermi} limits improve (roughly $F_{U.L.}\propto1/\sqrt{T_{obs}}$) but, due to the temporal evolution of the cascade signal, the echo flux also decreases.
As it is described in Sec.  \ref{sect:crpropa} we used, as primary VHE spectrum, a log-parabola up to 10 TeV. We also tested the possibility that $E_{max}$ might be larger ($E_{max}=50$ TeV) and how this affects our results. Since the spectrum is curved, at the largest energies the flux is very low. For this reason, although we see that moving from $E_{max}=10$ TeV to $E_{max}=50$ TeV the level of cascade increases especially at $E>50$ GeV, the overall cascade flux does not change dramatically and our main conclusion remains unchanged.

One of the reasons why, despite the very promising GRB, the IGMF remains unconstrained can be understood from Eq. \ref{sim_cascade}: the amount of cascade flux is proportional to the GRB time activity in the VHE band. We would need the activity to be at least a factor of 5 larger (namely $\Delta T_{activity}>25$ hours)  in order to exclude IGMF strengths larger than $10^{-20}$ G for $T_{obs}>9$ months. In this regard, we note the reported detection at VHE $\gamma$-rays from the afterglow of GRB 190829A by the H.E.S.S. Collaboration \cite{HESS21}. In this case, the estimated power law index of the intrinsic spectrum is again around $\sim2$, while the redshift is considerably lower ($z=0.0785$) than for GRB 190114C. But the time activity in the VHE band measured by H.E.S.S. is about 51 hours, more than a factor of 10 larger than the one of GRB 190114C, at a similar flux level. To test whether GRB 190829A could be a better target for IGMF studies we repeated the same procedure using a power law with index 2 but adding an exponential cutoff at 4 TeV (the maximum estimated energy in the VHE spectrum) as the primary spectrum. Due to the low redshift, the cascade SED in the energy range 0.1---100 GeV and for $B=10^{-20}$~G, after 1 month of observation time is more than 4 orders of magnitude lower than the \textit{Fermi}-LAT upper limits\footnote{We note that the recently detected GRB 221909A \cite{SwiftGCN} at $z\sim0.15$ \cite{VLTGCN} could be an ideal target for these studies. This is also thanks to the maximum photon energy detected by the LHAASO Collaboration \cite{LHAASOGCN} ($E_{max}\simeq18$ TeV). However, the lack of a published VHE spectrum and a reliable spectrum in the GeV domain \cite{Fermi-bti} make a correct estimation of the cascade echo flux impossible at the moment.} 

Back to GRB 190114C, from Fig. \ref{lightcurve_logpar} we see that the \textit{Fermi}-LAT upper limits decrease faster than the predicted cascade flux with the observation time. To verify whether for large observation times the GeV upper limits might be lower than the predicted cascade flux, we simulated the \textit{Fermi}-LAT sensitivity as a function of the observation time. Consequently, we used the same instrument response functions and diffuse models and re-scaled our 24 months exposure map to various times between 20 days and 25 years. We also assumed again a power law with spectral index 2, requiring at least a $2\sigma$ detection and 3 counts above 1 GeV. Finally, we compared the \textit{Fermi}-LAT sensitivity with the cascade light curve for $B=8\times10^{-21}$ G (the case in which we have the largest cascade flux within the first 2 years of observation time) extrapolated up to $T_{obs}=10^{4}$ d $\simeq27.4$ yr.

\begin{figure}[t]
\begin{center}
\includegraphics[width=\hsize]{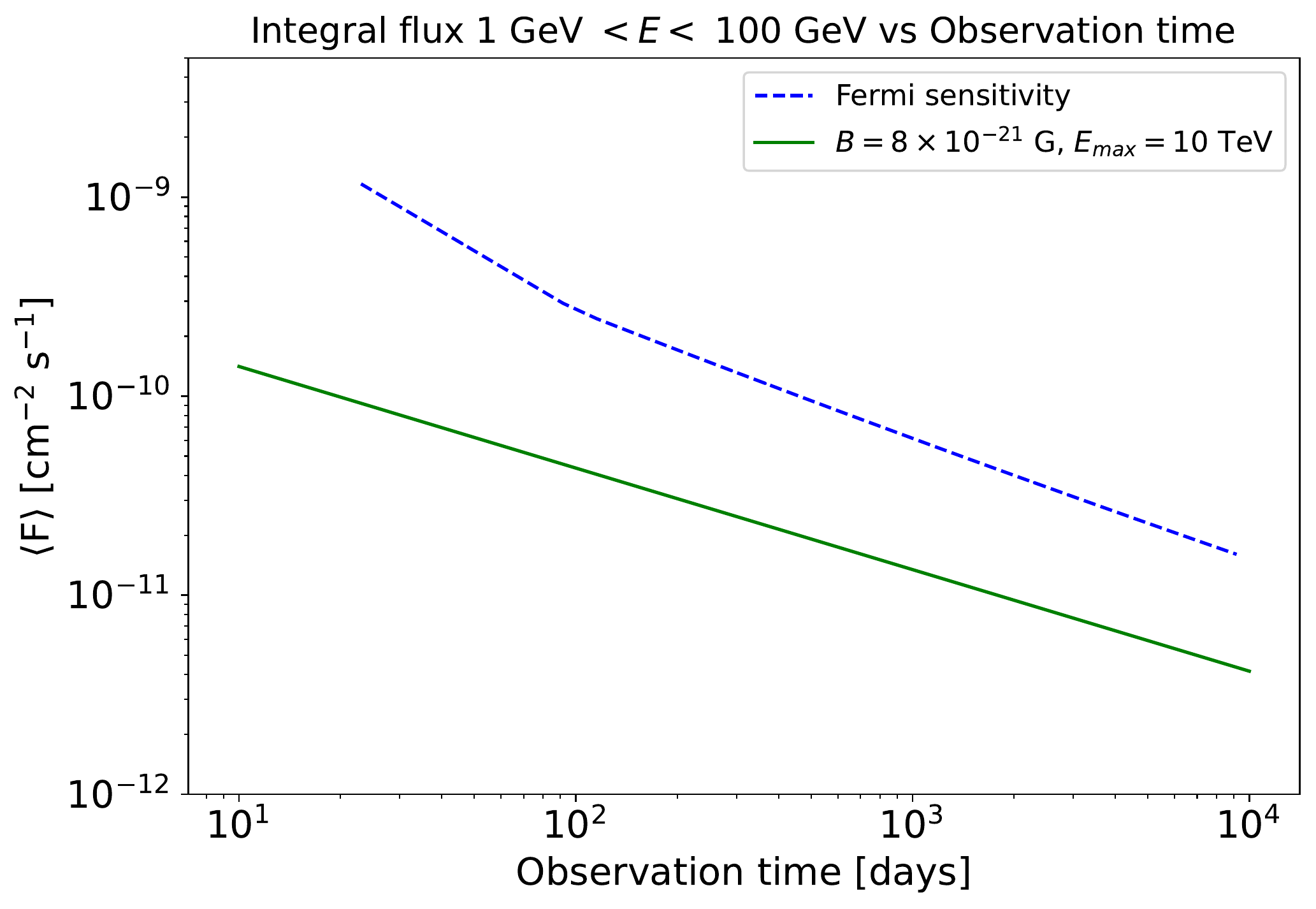}
\end{center}
\caption{\small \textit{Fermi}-LAT sensitivity (95\% confidence level) in the energy range $E=$ 1---100 GeV as a function of the observation time and simulated cascade lightcurve for $B=8\times10^{-21}$ G and $E_{max}$=10 TeV in the same energy band extrapolated up to $T_{obs}=10^4$ d. The slope change for times shorter than $80$ days is caused by the absence of at least 3 counts, requiring larger fluxes.}
\label{extrapolation}
\end{figure}

As we can see from Fig. \ref{extrapolation} from roughly $T_{obs}\simeq150$ d the sensitivity and the cascade lightcurve start to have the same slope. For this reason, there is no chance that the two curves can cross for a finite observation time.

Another test we performed concerns the \mbox{\emph{Fermi}-LAT} PSF: in Eq. \ref{sim_cascade} the cascade SED and lightcurve are calculated counting, in the simulations, the cascade photons within $\theta_{PSF}$. However, due to the deflection of the pairs, the cascade emission is also extended. As a consequence it might be possible that by increasing the angular extension used to compute the cascade SED and lightcurve, the level of cascade flux could increase. On the other hand, the \emph{Fermi}-LAT analysis should be changed accordingly because the morphological model assumed in the analysis described in the previous section is point-like. To verify this hypothesis we produced the angular distribution of the cascade in the first 24 months after the GRB: in this time range all the cascade photons are within the PSF of the instrument for each IGMF strength tested, therefore our result does not depend on the limited $\theta _{PSF}$ and no extension is expected.

As described in the introduction, two previous papers report different results. While in \cite{Wang20} the authors were able to calculate a lower limit on the IGMF strength, in \cite{Dzhatdoev20} no IGMF strengths can be constrained. In \cite{Wang20} the authors comment that this discrepancy can be due to the fact that Dzhatdoev et al. did not extrapolate the VHE flux up the first 6 seconds after the burst. This, of course, decreases significantly the cascade power. Although this is a crucial point, we find that even considering the extrapolation of the VHE flux up to $T_0+6$ s, no IGMF limits can be placed with this GRB. There is an important difference between our procedure and the ones adopted in \cite{Wang20} and \cite{Dzhatdoev20}: we chose, as the primary VHE spectrum, the one derived from the multiwavelength SED model published by the MAGIC Collaboration \cite{MAGIC19Model}. In this way our treatment is model dependent but, given the log-parabola shape (Eq. \ref{eqn:logpar}), the VHE flux at the highest energies is lower than the one we would have had choosing as primary spectrum a simple power law such as in \cite{Wang20} and \cite{Dzhatdoev20}. In this way, our choice is more conservative because the cascade power is lower. 
Furthermore, such a model justifies our extrapolation to earlier times as a reliable assumption: the fast cooling of the electrons likely implies that radiative losses start at the beginning of the afterglow, also shifting the peak of the synchrotron self-Compton component to lower energies -- thus the GRB would presumably exhibit harder spectra at earlier times \cite{MAGIC19Model}
In spite of this crucial difference, the cascade flux that we inferred is still lower than the reported one in the two cited papers and we cannot reproduce their results.      

We performed this study assuming that the only mechanism through which the electron-positron pairs lose energy is IC. An alternative competing energy loss mechanism to IC is through beam-plasma instabilities. The plasma instabilities were firstly proposed by Broderick et al. \cite{Borderick12} to explain the non-detection of the electromagnetic cascade in blazar SEDs at GeV energies, as well as the lack of extended emission. Many subsequent studies have attempted to quantify how the plasma instabilities can efficiently cool down the pairs (see e.g. \cite{Miniati14, Schlickeiser13, Sironi14, Chang16, Rafighi17, Vafin18, Vafin19, Batista19, Alawashra22}) compared to the IC process. But the results of these studies strongly depend on the assumptions used; the extreme contrast between parameters of the interacting components -- such as the huge difference between the densities of the electron beam and the background plasma -- make the impact of the instabilities on the development of a cascade almost impossible to evaluate. However, the instabilities might not be a problem for the specific case of a GRB: in order to develop themselves, the instabilities require a certain amount of time ($\sim$ 300 yr, \cite{Borderick12}). Since $\Delta T_{activity}$ is much lower than this characteristic time, the instabilities might not have enough time to develop \cite{Batista19}, making the studies of the IGMF by means of GRB and VHE flares of blazars robust.

\begin{acknowledgments}
The Fermi LAT Collaboration acknowledges generous ongoing support from a number of agencies and institutes that have supported both the development and the operation of the LAT as well as scientific data analysis. These include the National Aeronautics and Space Administration and the Department of Energy in the United States, the Commissariat \`{a} l'Energie Atomique and the Centre National de la Recherche Scientifique / Institut National de Physique Nucl\'{e}aire et de Physique des Particules in France, the Agenzia Spaziale Italiana and the Istituto Nazionale di Fisica Nucleare in Italy, the Ministry of Education, Culture, Sports, Science and Technology (MEXT), High Energy Accelerator Research Organization (KEK) and Japan Aerospace Exploration Agency (JAXA) in Japan, and the K. A. Wallenberg Foundation, the Swedish Research Council and the Swedish National Space Board in Sweden. Additional support for science analysis during the operations phase from the following agencies is also gratefully acknowledged: the Istituto Nazionale di Astrofisica in Italy and the Centre National d'Etudes Spatiales in France. This work performed in part under DOE Contract DE-AC02-76SF00515. P.V. acknowledges support from NASA grant NNM11AA01A.\\

\end{acknowledgments}



\bibliography{sample.bib}
\end{document}